\documentclass[journal]{IEEEtran}
\usepackage{amsthm}
\usepackage{amsmath}
\usepackage{amssymb}
\usepackage{latexsym}
\usepackage{bbm}
\usepackage[dvips]{graphicx}
\usepackage{enumerate}
\usepackage{verbatim}
\usepackage{amsfonts}
\usepackage{epsfig}
\usepackage{tikz}
\usetikzlibrary{fit}
\usetikzlibrary{shapes}
\usepackage{url}
\usepackage[lofdepth,lotdepth,caption=false]{subfig}

\usepackage{color}
\usepackage{epstopdf}
\usepackage{breqn}

\begin{document}
%
\title{Layered Coding of Hidden Markov Sources}
%
%
%

\author{Mehdi~Salehifar,~\IEEEmembership{Member,~IEEE,}
        Tejaswi~Nanjundaswamy,~\IEEEmembership{Member,~IEEE,}
        and~Kenneth~Rose,~\IEEEmembership{Fellow,~IEEE}
}

%
%

\markboth{}%
{}
%



\maketitle

\begin{abstract}
The paper studies optimal coding of hidden Markov sources (HMS), which represent a broad class of practical sources obtained through noisy acquisition  processes, beside their explicit modeling use in  speech processing and recognition, image understanding and sensor networks. A new fundamental source coding approach for HMS is proposed, based on tracking an estimate of the state probability distribution, and is shown to be optimal. Practical encoder and decoder schemes that leverage the main concepts are introduced. An iterative approach is developed for optimizing the system. It also focuses on a significant extension of the optimal HMS quantization paradigm. It proposes a new approach for scalable coding of HMS which accounts for all the available information while coding a given layer.  Simulation results confirm that these approaches significantly reduce the reconstructed distortion and substantially outperform  existing techniques.

\end{abstract}


%
\IEEEpeerreviewmaketitle

\section{Introduction}
\IEEEPARstart{T}{he} Hidden Markov model (HMM) is a discrete-time finite state Markov chain observed through a memoryless channel. The random process consisting of the sequence of observations is referred to as a hidden Markov source (HMS). Markov chains are common models for information sources with memory and memoryless channel is among the simplest communication models. HMMs are widely used in image understanding and speech recognition \cite{rabinerlevinson}, source coding \cite{gersho1992vector}, communications, information theory, economics, robotics, computer vision and several other disciplines. Note that most signals modeled as Markov process are in fact captured by imperfect sensors and are hence contaminated with noise, i.e., the resulting sequence is an HMS.
Motivated by its modeling capability of practical sources with memory,  we consider optimal quantization of the HMS. One conventional approach to design quantizers for the HMS is to employ predictive coding techniques (such as DPCM) to exploit time correlations, followed by standard scalar quantization of the residual. However, any direct prediction method cannot fully exploit the structure of the source process, i.e., the underlying Markov process. Indeed, even if the underlying process is first order Markov (depends on the past only through the previous sample), the HMS is not Markov, i.e, all prior samples are needed to optimally predict the current sample. But a naive infinite order prediction is clearly of impractical complexity. An alternative to predictive coding is encoding by a finite-state quantizer (see eg. \cite{foster1985finite,dunham1985algorithm}). A finite-state quantizer is a finite-state machine used for data compression: each successive source sample is quantized using a quantizer code book that depends on the encoder state. The current encoder state and the codeword obtained by quantization then determine the next encoder state, i.e., the next quantizer codebook. In  \cite{switch}, the authors optimized the finite-state quantizer for a given HMS with known parameters, and analyzed its steady state performance. While a finite state quantizer is, in general, an effective approach, it does not fully exploit the underlying structure of the HMS. In other words, the finite state quantizer is agnostic to the {\em source state}, although it updates the {\em encoder state} based on the HMS observations. This subtle but important shortcoming motivates a scheme that {\em explicitly} considers the source states in the quantization process, which is the approach pursued in our previous work  \cite{HMM}.

The fundamental approach is to exploit all available information on the source state, i.e., the probability distribution over the states of the underlying Markov chain. An important distinction of this paradigm from prior work, and specifically from the finite-state quantizer, is that the state probability distribution captures all available information and depends on the entire history of observations. Hence, with each observation both encoder and decoder refine the estimate of state probability distribution and correspondingly update the coding rule. The fundamental approach is therefore optimal. We then specialize to a practical ``codebook switching'' variant and propose an algorithm that optimizes the decision rule, i.e., we optimize the codebook used for quantization based on the estimate of state probabilities.

Also we focused on a significant extension of the optimal HMS quantization paradigm. Advances in internet and communication technologies, have created an extremely heterogeneous network scenario with data consumption devices of highly diverse decoding and display capabilities, all accessing the same content over networks of time varying bandwidth and latency. Thus it is necessary for coding and transmission systems to provide a scalable bitstream that allows decoding at a variety of bit rates (and corresponding levels of quality), where the lower rate information streams are embedded within the higher rate bitstreams in a manner that minimizes redundancy. That is, we need to generate layered bitstreams, wherein a base layer provides a coarse quality reconstruction and successive layers refine the quality, incrementally.  Scalable coding with two quality levels, transmits at rate $R_{12}$ to both decoders and at rate $R_2$ to only the second decoder.

	The commonly employed technique for scalable coding of HMS, completely neglects the hidden states and employs a predictive coding approach such as DPCM (i.e., assumes a simple Markov model) at the base layer, and at the enhancement layer, the base layer reconstruction error is compressed and transmitted. That is, the base layer reconstruction is used as an estimate for the original signal, and the estimation error is compressed in the enhancement layer. A review of scalable coding techniques for audio and video signals can be found in \cite{svc,svc2}. An estimation theoretically optimal scalable predictive coding technique was proposed in \cite{PredictiveSC}, which accounts for all the information available from the current base layer and past reconstructed samples to generate an optimal estimate at the enhancement layer. In another work of our lab \cite{sca_HMM,phd}, we proposed a novel scalable coding technique for HMS, which accounts for all the available information while coding a given layer. At the base layer, we exploit all the available information by employing our previously proposed technique of tracking the state probability distribution at the encoder and the decoder, and using it to update the quantizers for encoding the current observation. At the enhancement layer, we again track the state probability distribution at the encoder and the decoder, but using the higher quality enhancement layer reconstruction for a better estimate, and then the enhancement layer quantizer is adapted to the interval determined by base layer quantization so as to enable full exploitation of all available information. 

Beyond what has been done before in our previous works, in this paper we tackle the new problems of the unreliable networks, delayed scalable coding, and designing optimal coding for audio's LSF attribute. 

Multimedia transmission over networks enables a wide range of applications such as online radio, TV and high-definition teleconferencing. These applications are often suffering by the problem of unreliable networking conditions, which leads to the loss of data. Packet loss concealment is an important tool to solve this issue. The packet loss concealment objective is to exploit all available information to approximate the lost packet while maintaining smooth transition with neighboring packets. In this paper we propose an approach, which the state probability distribution tracking unit, considers two scenarios of either having a packet loss or not, and based on that it updates the probability distribution and design the other parameters.  

An important feature of scalable coding is that the base layer can be decoded independently of enhancement layers, which allows the enhancement layer coder potential access to information about future base layer reconstruction, at a given coding latency relative to the base layer. In this paper we also consider means to exploit such future information, in addition to the current base layer and prior enhancement layer information, in a scheme that complements the above scalable framework, to further refine the enhancement layer probability distribution, and thereby achieve considerable performance gains on top of the non-delayed approach.

Finally, we tackle the problem of designing coding approach for LSF data in this paper. Linear predictive coding (LPC) is a tool used mostly in speech processing for representing the spectral envelope of a digital signal of speech in compressed form. It is one of the most useful methods for encoding good quality speech at a low bit rate and provides extremely accurate estimates of speech parameters. Transmission of the filter coefficients directly is undesirable, since they are very sensitive to errors. Line spectral frequencies (LSF) are used to represent LPC for transmission over a channel. In this paper we design the best coding approach for LSF coefficients.

The rest of this paper is organized as follows: In Section II, we overview background information. In Section III and IV, the proposed methods are described. Experimental results are presented in Section V and concluding remarks are in Section VI.


\begin{section}{Preliminaries }
	
	The hidden Markov model drives its name from two defining properties: First it assumes that the observation at time $t$ is generated based on a state which is hidden from observer. Secondly these state sequences are related through a Markov process in which the current state's dependence on the past is completely determined by the immediately preceding state.
	
	 HMMs have become more  popular in various important applications. 
	 HMMs are widely used in image understanding and speech recognition \cite{rabinerlevinson, Juang14,Juang15,Juang16,Juang17,Juang18}, source coding \cite{gersho1992vector}, communications, information theory, economics, robotics, computer vision and several other disciplines. Note that most signals modeled as Markov processes are usually captured by imperfect sensors and are hence contaminated with noise, i.e., the resulting sequence is in fact an HMS. HMS is a special case of the broader family of multivariate Markov sources, which has been a focus of recent research, notably in the context of parameter estimation  \cite{forward_bi,causal_H_bi}. 
	
	  Following we will specify the formal definition and the most basic problems associated with the  HMS.
	\begin{subsection}{Definition of Hidden Markov Sources}
		A  hidden Markov source (HMS) is the output of a hidden Markov model (HMM) determined by five parameter sets (see eg., \cite{rabiner}):
		\begin{enumerate}
			\item The number of states in the model $N$. We denote the set of all possible states as $S = \{S_1, S_2, \ldots, S_N\}$, and the state at time $t$ as $q_t$.
			\item The state transition probability distribution $A = \{a_{ij}\}$, where $a_{ij}= P[q_{t+1} = S_j|q_t = S_i], 1 \leq i,j \leq N$.
			\item Observation symbols, in discrete case we denote the symbols as   \quad $V= \{ v_1, v_2, \cdots, v_M\}$, where $M$ is the number of distinct observation symbols per state.
			\item The observation (emission) probability density function (pdf) in state $j$, $g_j(x)$, in the continuous case, and the observation (emission) probability mass function (pmf), $B = \{b_j(v_k)\}$, where $b_j(v_k)= p[O_t=v_k|q_t=S_j]$ in the discrete case. $O_t$ denotes the source output at time $t$.
			\item The initial state distributions $\pi = \{\pi_i\}$ where $\pi_i =P[q_1=S_i], 1 \leq i \leq N$.
		\end{enumerate}
		
		\begin{figure}[t]
			\centering
			\centerline{\includegraphics[width=.5\textwidth]{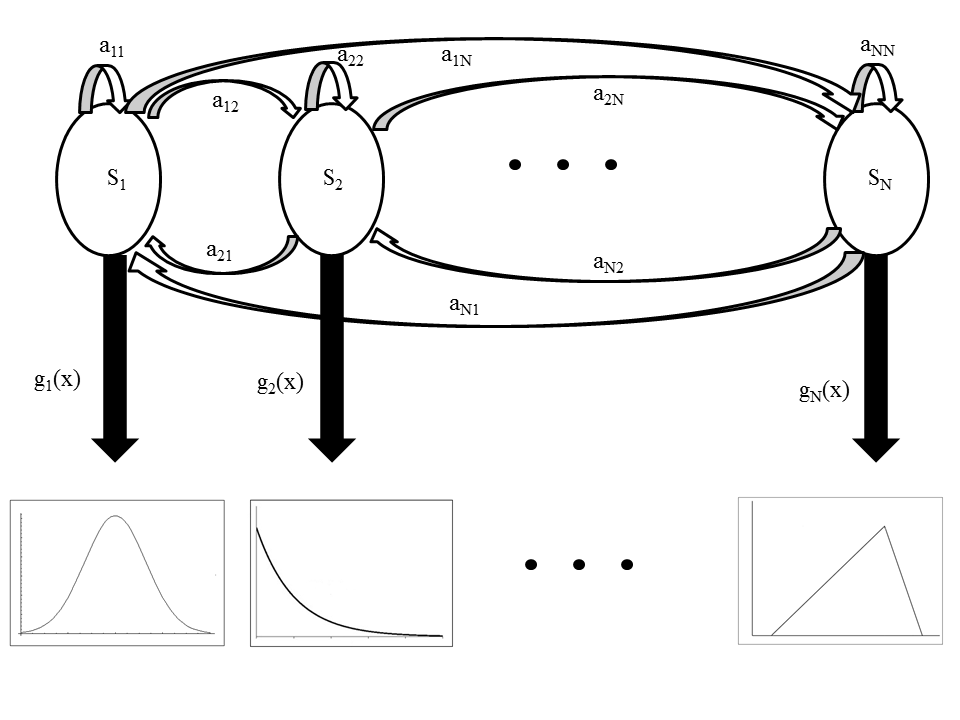}}
			
			\caption{A continuous emission hidden Markov source with $N$ states.}
			\label{fig:HMM_gen}
		\end{figure}

		Fig.\ \ref{fig:HMM_gen} depicts an example of a continuous hidden Markov source with $N$ states.
		In hidden Markov sources we have no access to the state of the source, only to the emitted observation, $O_n$.
		
		For convenience we use the notation $\lambda = (A,B,\pi)$ to indicate the parameter set.
		
	\end{subsection}
	\begin{subsection}{Three Basic Problems}
		To use the hidden Markov model in real-world applications, there are typically three very fundamental problems that need to be solved (see eg., \cite{rabiner}):
		\begin{enumerate}
			\item Given the model $\lambda = (A,B,\pi)$ what is the probability of generating the observation sequence $O_{1\rightarrow T} = O_1 O_2, \ldots, O_T$, i.e.,  $P(O_{1\rightarrow T}|\lambda)$.
			\item Given the observation sequence $O_{1\rightarrow T} $ and the model $\lambda = (A,B,\pi)$ what is the most probable underlying state sequence $q_1^T = q_1 q_2, \ldots, q_T$.
			\item Given the observation sequence $O_{1\rightarrow T}$ how to estimate the HMM parameters  $\lambda = (A,B,\pi)$.
		\end{enumerate}
		
		\textbf{Solution to the First Problem: Evaluation Problem }
		
			We want to calculate the probability of the observation $O_{1\rightarrow T} = O_1 O_2, \ldots, O_T$ given the model $\lambda$, i.e. $P(O_{1\rightarrow T}|\lambda)$. 
%
			  An efficient way to solve the problem is by using the so-called Forward and Backward procedure:
			 Let us define the Forward variable $\alpha_t(i)$ as :
			 
			 \begin{align}
			 \alpha_{t}(i) = P(O_{1\rightarrow t}, q_{t}=S_i),
			 \end{align}
			 i.e., the joint probability of emitting the observed source sequence  $O_1^{t}$ up to time $t$ and that the state at time $t$ is $S_i$.
			 The forward variables can be computed recursively:
			 \begin{enumerate}
			 	\item $\alpha_1(i)= \pi_ib_i(O_1)$ for $ 1\leq i\leq N$
			 	\item $\alpha_{k+1}(j)=[\sum_{i=1}^N \alpha_k(i)a_{ij}]b_j(O_{k+1})$
			 \end{enumerate}
			 Finally probability of the sequence observation $O_1^T$ is :
			 \begin{align}P(O_{1\rightarrow T}|\lambda)= \sum_{i=1}^N \alpha_T(i) \end{align}
			 In a similar manner we could define Backward variable $\beta_t(i)$:
			 \begin{align}
			 \beta_t(i)= P[O_{(t+1)\rightarrow T}| q_t=S_i]
			 \end{align}
			 i.e., the probability of emitting symbols $O_{t+1} , \ldots , O_T$, given that the source was in state $i$ at time $t$. These backward variables can be updated recursively as follows:
			 \begin{enumerate}
			 	\item $\beta_T(i)=1$,  for $ 1\leq i\leq N$
			 	\item $\beta_t(i)= \sum_{j=1}^N a_{ij}b_j(O_{t+1})\beta_{t+1}(j)$,	for $  1\leq t\leq T-1, 1\leq i\leq N$
			 \end{enumerate}
			 
			  Finally probability of the sequence observation $O_1^T$ is :
			  \begin{align} P(O_{1\rightarrow T}|\lambda)= \sum_{i=1}^N \beta_1(i) b_{S_i}(O_1) \end{align}

		\textbf{Solution to the Second Problem: Optimal Path}
		
			There are several ways to define the optimal state. One way is to choose the state $q_t$ individually such that it is most likely. To implement this solution let us define the variable :
			\begin{align} \gamma_t(i) = P(q_t = S_i|O_{1\rightarrow T})\end{align}  i.e., the probability to be at state $i$ at time $t$ given the observation sequence $O_1^T$. Using forward and backward variable:
			\begin{align}
				 \gamma_t(i) &= \frac{\alpha_t(i)\beta_t(i)}{P(O_{1\rightarrow T})} \nonumber\\
				 & = \frac{\alpha_t(i)\beta_t(i)}{\sum_{i=1}^N \alpha_t(i)\beta_t(i)}
			\end{align}
			Using $\gamma_t(i)$ we can solve for individually optimal state $q_t$ at time $t$:
			\begin{align}q_t = argmax (\gamma_t(i))\end{align} for all $t$.
			
			However if we are interested  in optimal sequence of states then  the viterbi algorithm could be used to find the optimal sequence path.
		
		\textbf{Solution to the Third Problem: Parameter Training}
		
			The third problem is the most difficult HMM problem and there is no way to analytically solve this problem. However we can choose $\lambda = (A,B,\pi)$, such that $P(O_1^T|\lambda)$ be locally maximized using an iterative procedure such as Baum-Welch (or any other EM) method as explained in \cite{rabinerlevinson}.

			\end{subsection}	
		
\end{section}
\begin{section}{Optimal Coding of HMS}
\label{sec:pagestyle}

One important property of the HMS is that the state at time $t-1$, captures all past information relevant to the emission of  the next source symbol. Specifically $ P[O_t= v_k| q_{t-1}=S_j, O_1^{t-1}]= P[O_t= v_k| q_{t-1}=S_j]$ which implies that all  observations until time $t-1$ provide no additional information on  the next  source symbol, beyond what is obtained by  the state of the Markov chain at time $t-1$. Note further that we can not know with certainty the state of the HMS. Based on this fact the fundamental paradigm for achieving optimal coding is to track the state probability distribution of the HMS. In this approach, each output of the encoder (quantized observation) is sent into a unit called the \emph{state probability distribution tracking function}. This unit estimates the state probability distribution, i.e., probabilities of the Markov chain being in each of the states, denoted by  $\underline{\hat{p}}$. These probabilities are used in the \emph{encoder adapter unit} to redesign the encoder optimally for the next input sample. Fig.\ \ref{fig:gen_enc} shows the fundamental approach for  the optimal HMS encoder.

\begin{figure}[t]
	\centering
	\centerline{\includegraphics[width=.5\textwidth]{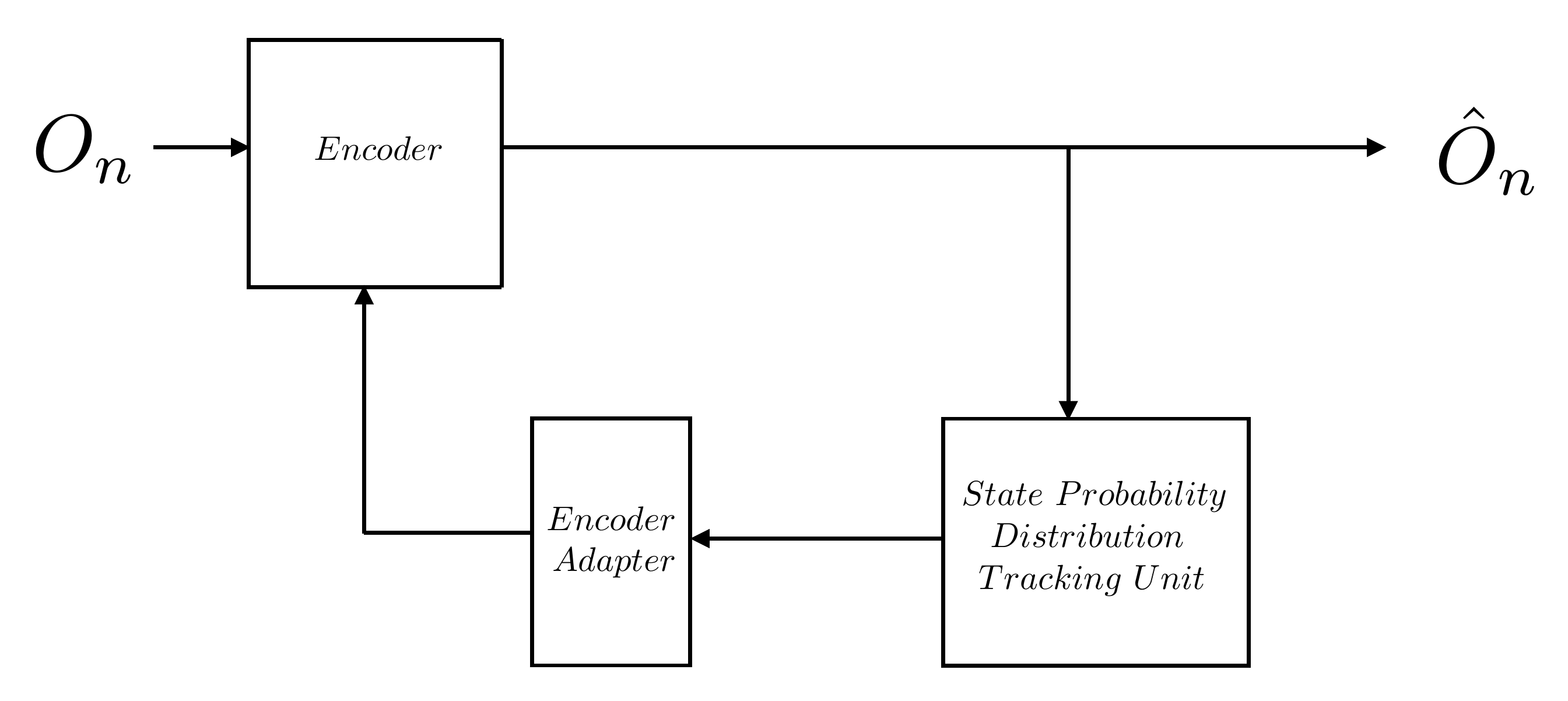}}
	\caption{Fundamental approach to optimal encoding of a hidden Markov source.}
	\label{fig:gen_enc}
\end{figure}



Our objective is to design the Encoder,  the state probability distribution tracking function and the Encoder adapter, given a training set of samples, so as to minimize the average reconstruction distortion at a given encoding rate. We first describe tracking of the state probability distribution, then we describe fundamental design approach, then proceed to practical encoder design approach.	Finally we describe the approach under lossy network condition.

	\subsection{State Probability Distribution Tracking}
	\label{sec:spdtrack}
	We first estimate the HMS parameters using a training set, for which we follow the procedure explained earlier. Using forward variable at time $t-1$ as
	\begin{align}
	\alpha_{t-1}(i) = P(O_{1 \rightarrow(t-1)}, q_{t-1}=S_i),
	\end{align}
	i.e., the joint probability of emitting the observed source sequence up to time $t-1$, $O_{1\rightarrow (t-1)}$ and that the state at time $t-1$ is $S_i$.
	
	We can then obtain probability of being in state $i$ at time $t - 1$, using Bayes' rule,
	\begin{align}
	P[q_{t-1} = S_i| O_{1\rightarrow(t-1)}] &= \frac{P(q_{t-1}=S_i, O_{1\rightarrow(t-1)})}{P(O_{1\rightarrow(t-1)})} \nonumber\\
	&= \frac{P(q_{t-1}=S_i, O_{1\rightarrow(t-1)})}{\sum_{j=1}^N P(q_{t-1}=S_j, O_{1\rightarrow(t-1)})} \nonumber\\
	&= \frac{\alpha_{t-1}(i)}{\sum_{j=1}^N \alpha_{t-1}(j)}
	\end{align}
	Using these we can compute the probability of being in state $i$ at time $t$, $p_{(t,i)}$, given the observation sequence up to time $t-1$ as: 
	\begin{align}
	p_{(t,i)} &= P[q_t= S_i| O_{1\rightarrow(t-1)}] \nonumber\\
	&= \sum_{j=1}^N P[q_{t-1}= S_j| O_{1\rightarrow(t-1)}]a_{ij}
	\end{align}
	The state probability distribution at time $t$, $\underline{p}_{(t)}$, is defined as,
	\begin{align}
	\underline{p}_{(t)} &= \{p_{(t,1)}, \ldots, p_{(t,N)}\} \nonumber\\
	&= \{P[q_t= S_1| O_{1\rightarrow(t-1)}],\cdots,P[q_t= S_{N}| O_{1\rightarrow(t-1)}]\}
	\end{align}

	\begin{figure*}[t]
		\centering
		\centerline{\includegraphics[width=1\linewidth]{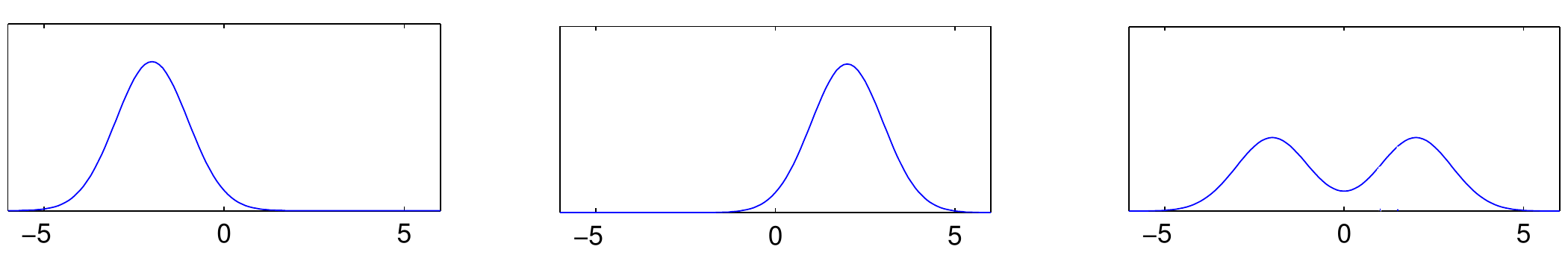}}
		\caption {Example of observation pdf at states 1 and 2, $g_1(x)$ and $g_2(x)$ respectively, and the overall observation pdf  given $\hat{p}_{(t)}=\{0.5,0.5\}$,  $P(O_t|\hat{p}_{(t)}=\{0.5,0.5\})$}
		\label{fig:base}
	\end{figure*}
\subsection{Fundamental Design Approach}
The shortcoming of tracking the state probability distribution based on original source symbols is that the decoder cannot exactly mimic the encoder. To avoid this problem the state probability distribution  tracking unit must track the states probability distribution, based on the output of the encoder.

In fundamental approach, state probability distribution tracking unit finds, $\underline{\hat{p}}$, using the reconstructed observation samples $\hat{O}$, as
	\begin{align}
	\underline{\hat{p}}_{(t)} &= \{\hat{p}_{(t,1)},\ldots,\hat{p}_{(t,N)}\} \nonumber\\
	&= \{P[q_t= S_1| \hat{O}_{1\rightarrow(t-1)}],\ldots,P[q_t= S_{N}| \hat{O}_{1\rightarrow(t-1)}]\}
\end{align}
		Given $\underline{\hat{p}}_{(t)}$, the best estimate for the source distribution is:
	\begin{align}
	\sum_{j=1}^{N}\hat{p}_{(t,j)}g_j(x)
	\label{eqn:base}
	\end{align}

	Fig. \ref{fig:base} depicts this notion.
	
	In this approach, each output of the encoder (quantized observation) is sent into the state probability distribution tracking function. This unit estimates the state probability distribution, i.e., $\underline{\hat{p}}$. These probabilities are used in the encoder adapter unit to design a quantizer for the pdf  in Eq.~\ref{eqn:base} for each source output from the ground up. Finally Encoder encode the symbols. 
	

\subsection{Practical Encoder Design}

The fundamental framework requires the encoder be redesigned for each observed sample, which entails high complexity. Therefore, to propose a practical approximation with both low complexity and low memory requirement, we restrict the encoder adapter unit to select a codebook from a pre-designed set of codebooks and use it as an initialization. To facilitate this operation, we introduce a module called the `next quantizer selector', which decides the codebook to be used for initialization of the codebooks of the next sample, based on the output of the state probability distribution tracking function. Fig.\ \ref{fig:encoder} shows the proposed practical encoding scheme. 

 For a set of $T$ representative $\underline{p}$, we find the best codebook using Lloyd's (or other) algorithm offline. 
 \begin{align*}
	C_n=\{x_{n1},\ldots,x_{n(2^R)}\}  \quad \text{for} \quad n=1,2,\ldots,T
\end{align*}
 Then for a given $\underline{\hat{p}}_{(t)}$ in the process of encoding or decoding, we find the closest representative  $\underline{p}$ from the set of $T$. Finally using the codebook of the closest representative as an initialization, we run one iteration of the Lloyd's algorithm, to find the codebook to be used for current sample.

\subsection{Packet Loss Concealment for HMS}
	Audio and video transmission over networks enables a wide range of applications such as multimedia streaming, online radio, TV and high-definition teleconferencing. These applications are often plagued by the problem of unreliable networking conditions, which leads to intermittent loss of data. This problem is severe, and the packet loss rate in Internet communications, for example, may reach 20\%. Clearly, robustness to packet loss is a crucial requirement. Packet loss concealment forms a crucial tool amongst the various strategies used to mitigate this issue. The packet loss concealment objective is to exploit all available information to approximate the lost packet while maintaining smooth transition with neighboring packets.
	
	Let  $p_{loss}$ be the probability of the network's packet loss before it reaches the decoder. In another words, by probability $p_{loss}$, the encoded packet will not reach to the decoder, and gets lost. Then the decoder calculation of 
	\begin{align}
	\underline{\hat{p}}_{(t+1)} &= \{\hat{p}_{((t+1),1)},\ldots,\hat{p}_{((t+1),N)}\} \nonumber\\
	&= \{P[q_t= S_1| \hat{O}_{1\rightarrow t}],\ldots,P[q_t= S_{N}| \hat{O}_{1\rightarrow t}]\}
	\end{align}
	has two scenarios:
	\begin{itemize}
		\item If there is no packet loss at time $t$, the decoder would have access to $\hat{O}_{t}$ and would calculate the $\hat{p}_{((t+1),j),no-loss}$ as:
		\[ \hat{p}_{((t+1),j),no-loss}=C\sum_{i=1}^N \hat{p}_{(t,i)}a_{ij}b_i(\hat{O}_{t})\]
		\item If there is  packet loss at time $t$ then the decoder would not have access to $\hat{O}_{t}$ and would calculate the $\hat{p}_{((t+1),j),with-loss}$ as:
		\[ \hat{p}_{((t+1),j),with-loss}=\sum_{i=1}^N \hat{p}_{(t,i)}a_{ij}\]
	\end{itemize}
	Since the decoder with probability $1-p_{loss}$ would receive the packet and calculates the state probability as $\hat{p}_{((t+1),j),no-loss}$ and with probability $p_{loss}$ it would miss the packet and calculates the state probability as $\hat{p}_{((t+1),j),with-loss}$,the encoder's best estimate  of the decoder's probability distribution would be:
	\begin{align}
	\hat{p}_{((t+1),j),lossy}= &(1-p_{loss})\  \hat{p}_{((t+1),j),no-loss}\nonumber\\
	&+ p_{loss}\  \hat{p}_{((t+1),j),with-loss}
	\end{align}
	For the design procedure, we first estimate $\underline{\hat{p}}_{(t+1) ,lossy} $ for all $t$, later as before we vector quantize $p$ and design the  best codebook for each class. Also for each class we store the average of the class as the representative.  
	
	\begin{figure}[t]
	\centering
	\centerline{\includegraphics[width=.5\textwidth]{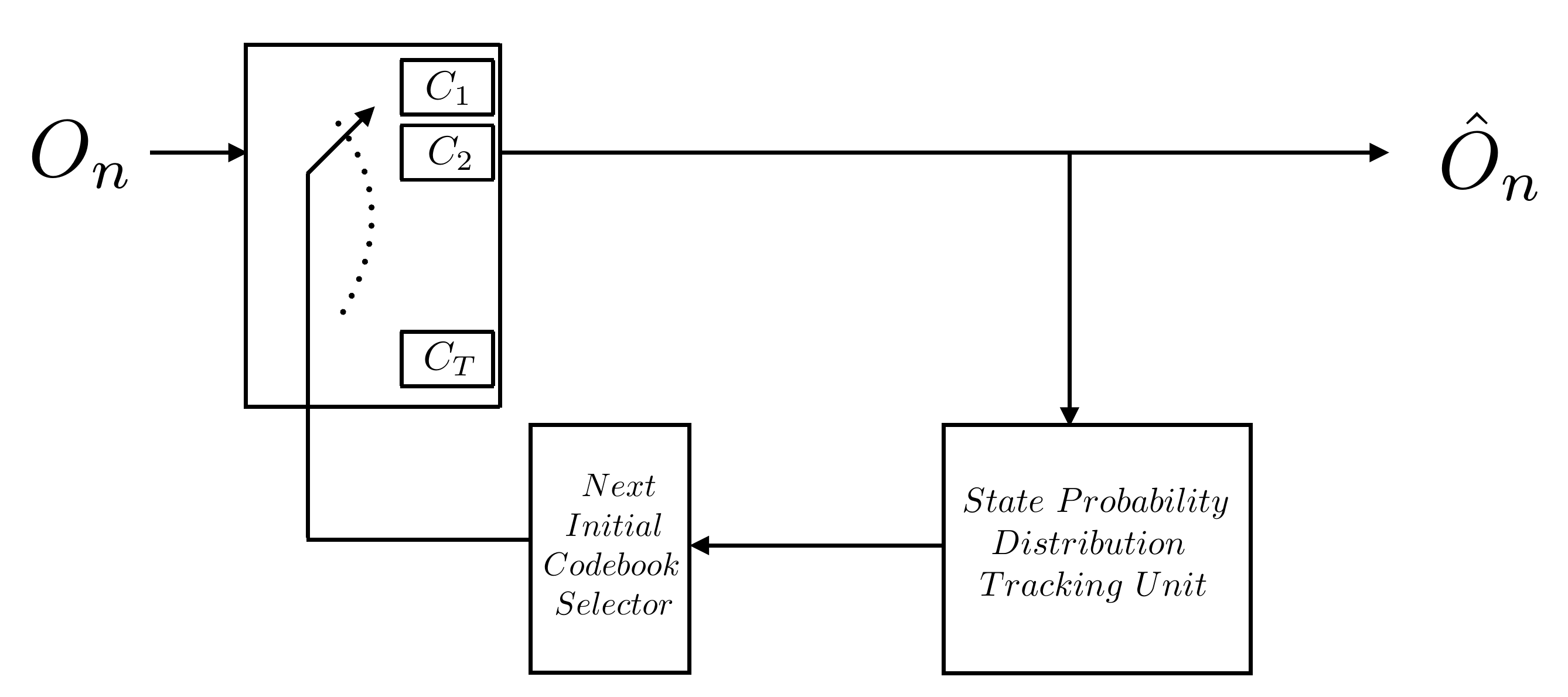}}
	\caption{Proposed practical encoder scheme.}
	\label{fig:encoder}
\end{figure}

	At the decoder side, depending on whether  we had packet loss at time $t$ or not, the decoder chooses one of the above scenarios to calculate the probability distribution at time $t+1$. 
	
	If there is no packet loss at time $t+1$ the decoder, as usual, finds the correct codebooks and decodes the received symbol. However if there is packet loss at time $t+1$ the decoder conceals it by substituting the decoded symbol by a weighted average of the class representatives, where the weights are given by  the probability distribution over states at time $t+1$.

	\begin{figure*}[t]
		\centering
		\centerline{\includegraphics[width=.9\linewidth]{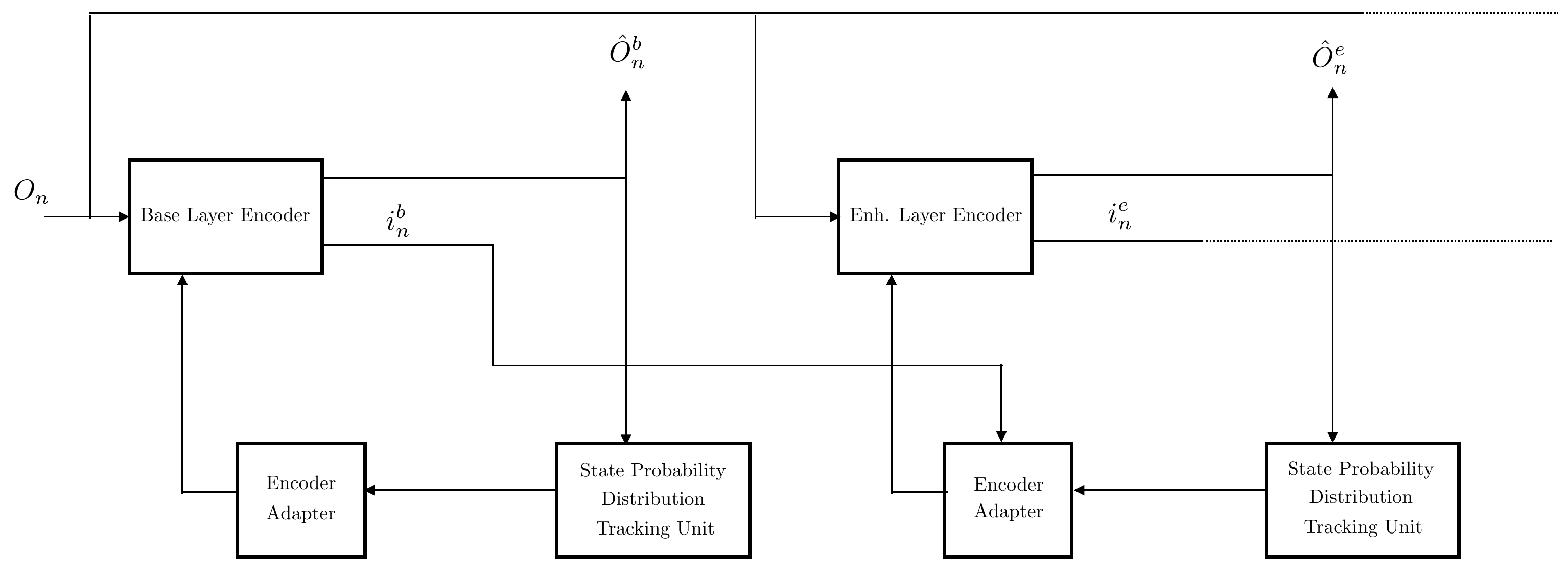}}
		
		\caption{Proposed encoder for the base and the enhancement layer.} 
		\label{fig:Enco_sca}
	\end{figure*}

\textbf{Note:} Before concluding this section it is important to  note that for the first symbol, we simply use a quantizer based on the assumption that the source symbol is from a fixed state (based on $\pi$) at both the encoder and decoder.
	
\end{section}

\begin{section}{Scalable Coding of HMS}

	For optimal scalable coding of HMS, we need to exploit all the available information while encoding at each layer.

	In this approach, each output of the encoder (quantized observation) at a given layer is sent into the state probability distribution tracking unit. This unit estimates the state probability distribution at each layer. 
	
	At the base layer the state probability distribution tracking unit finds, $\underline{\hat{p}^b}$, using base layer reconstructed observation samples $\hat{O}^b$, as
	\begin{align}
	\underline{\hat{p}}_{(t)}^b &= \{\hat{p}_{(t,1)}^b,\ldots,\hat{p}_{(t,N)}^b\} \nonumber\\
	&= \{P[q_t= S_1| \hat{O}_{1\rightarrow(t-1)}^b],\ldots,P[q_t= S_{N}| \hat{O}_{1\rightarrow(t-1)}^b]\}.
	\end{align}
	Finally using the approach explained in Section \ref{sec:pagestyle}, we design the best encoder.
	
	 At the enhancement layers, there is additional information of the quantization interval from the lower layers along with the state probability distribution. Thus the enhancement layer \emph{encoder adapter unit} combines both types of available information to redesign the encoder optimally for the next input sample. Fig.\ \ref{fig:Enco_sca} shows the proposed scalable encoder of HMS.
	
	\begin{figure*}[t]
		\centering
		\centerline{\includegraphics[width=.8\linewidth]{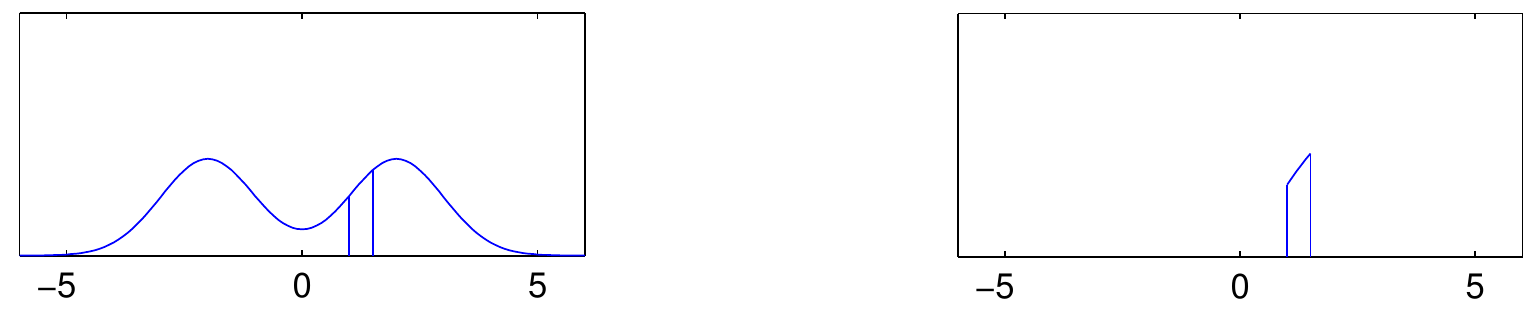}}
		\caption {Overall observation pdf in the enhancement layer given $\hat{p}_{(t)}=\{0.5,0.5\}$ and  the quantization interval information from the base layer ($I_b$),  $ P(O_t|\hat{p}_{(t)}=\{0.5,0.5\}, I_b)$.}
		\label{fig:enh}
	\end{figure*}
	
	
	Our objective is to design the  the encoder adapter for enhancement layes, given a training set of samples, so as to minimize the average reconstruction distortion at a given encoding rates for enhancements layers.

	\subsection{Enhancement Layer Design}
	The goal here is to design the best codebook for the enhancement layer based on:
	\begin{itemize}
		\item Quantization interval from the base layer, and
		\item State probability distribution based on the enhancement layer reconstructed observation samples $\hat{O}^e$.
	\end{itemize}

	The enhancement layer state probability distribution tracking unit finds $\underline{\hat{p}}^e_{(t)}$, similar to the base layer, but using enhancement layer reconstructed observation samples $\hat{O}^e$, as
	\begin{align}
	\underline{\hat{p}}_{(t)}^e &= \{\hat{p}^e_{(t,1)},\ldots,\hat{p}^e_{(t,N)}\} \nonumber\\
	&= \{P[q_t= S_1| \hat{O}_{1\rightarrow(t-1)}^e],\ldots,P[q_t= S_{N}| \hat{O}_{1\rightarrow(t-1)}^e]\}.
	\end{align}
	Given just $\underline{\hat{p}}^e_{(t)}$, the best estimate for the source distribution is:
	\begin{align}
	\sum_{j=1}^{N}\hat{p}_{(t,j)}^eg_j(x)
	\end{align}
	However, in combination with the quantization interval information from the base layer, the best estimate for the source distribution is:
	\begin{align}
	\sum_{j=1}^{N}\hat{p}_{(t,j)}^e\hat{g}_j(x)
	\end{align}

	where, $ \hat{g}_j(x) $ is the observation pdf in state $ j $  truncated and normalized to the interval determined by quantization at the base layer.   Fig. \ref{fig:enh} illustrates the best estimate for the source distribution using  interval information from the base layer.
	
	We design the quantizer for this pdf of the current sample, by using uniform codebook as an initialization and running one iteration of Lloyd's algorithm.
	
	Note that we use uniform codebook as an initialization for the enhancement layer quantizer design as observation pdf within the quantization interval of the base layer has lesser variations and is closer to uniform distribution. However, this assumption is not true for the base layer. Using the uniform codebook as the initialization also reduces the memory requirements of the encoder and the decoder. For the first symbol, we assume that the source symbol is from a fixed state (based on $\pi$) at both the encoder and decoder, and use this in combination with the quantization interval information from the base layer.

	Note that we can generalize our proposed approach to any number of enhancement layers, by combining the refined estimate of state probability distribution based on observation reconstruction of the given layer, with the quantization interval information from its lower layers.
	\subsection{Enhancement Layer Design With Access to Future Reconstruction of Base Layer}
	The goal here is to design the best codebook for the enhancement layer based on:
	\begin{itemize}
		\item Quantization interval from the base layer, and
		\item State probability distribution based on
		\begin{itemize}
			\item The enhancement layer reconstructed observation samples $\hat{O}^e$ till time $t-1$
			\item The base layer reconstructed observation samples $\hat{O}^b$ till time $t-1+L$, where $L$ shows the number of the future samples from base layer. 
		\end{itemize} 
		\end{itemize}
		
		The enhancement layer state probability distribution tracking unit finds $\underline{\hat{p}}^e_{(t)}$, similar to the base layer, but using enhancement layer reconstructed observation samples $\hat{O}^e$ till time $t-1$ as well as base layer reconstructed observation samples $\hat{O}^b$ till time $t-1+L$ , as
		\begin{align}
		\underline{\hat{p}}_{(t)}^e &= \{\hat{p}^e_{(t,1)},\ldots,\hat{p}^e_{(t,N)}\} \nonumber\\
		&= \{P[q_t= S_1| \hat{O}_{1\rightarrow(t-1)}^e,\hat{O}_{t\rightarrow(t-1+L)}^b],\ldots, \nonumber\\
		&~~~~P[q_t= S_{N}| \hat{O}_{1\rightarrow(t-1)}^e,\hat{O}_{t\rightarrow(t-1+L)}^b]\}
		\end{align}
		
		For finding $P[q_t= S_{i}| \hat{O}_{1\rightarrow(t-1)}^e,\hat{O}_{t\rightarrow(t-1+L)}^b]$, we use the forward-backward approach explained earlier. First we define the forward variable: 
		\begin{align}
			\alpha_{t}(i) = P(\hat{O}_{1\rightarrow(t-1)}^e,\hat{O}_{t}^b, q_{t}=S_i),
		\end{align}
		and the backward variable:
		\begin{align}
			\beta_t(i)= P[\hat{O}_{t+1\rightarrow(t-1+L)}^b| q_t=S_i]
		\end{align}
		and finally state probability is calculated as:
		\begin{align}
		P[q_t= S_{i}| \hat{O}_{1\rightarrow(t-1)}^e,\hat{O}_{t\rightarrow(t-1+L)}^b] = \frac{\alpha_t(i)\beta_t(i)}{\sum_{i=1}^N \alpha_t(i)\beta_t(i)}
		\end{align}
		for $i = 1,2, \cdots N$.
		
		Given just $\underline{\hat{p}}^e_{(t)}$, the best estimation for the source distribution is:
		\begin{align}
		\sum_{j=1}^{N}\hat{p}_{(t,j)}^eg_j(x)
		\end{align}
		However, in combination with the quantization interval information from the base layer, the best estimate for the source distribution is:
		\begin{align}
		\sum_{j=1}^{N}\hat{p}_{(t,j)}^e\hat{g}_j(x)
		\end{align}
		where, $ \hat{g}_j(x) $ is the observation pdf in state $ j $  truncated and normalized to the interval determined by quantization at the base layer.
		
		We design the quantizer for this pdf of the current sample, by using uniform codebook as an initialization and running one iteration of Lloyd's algorithm.
\end{section}

\begin{section}{Experimental Results}
\label{sec:typestyle}

\begin{figure}[t]
	\centering
	\centerline{\includegraphics[width=.5 \textwidth]{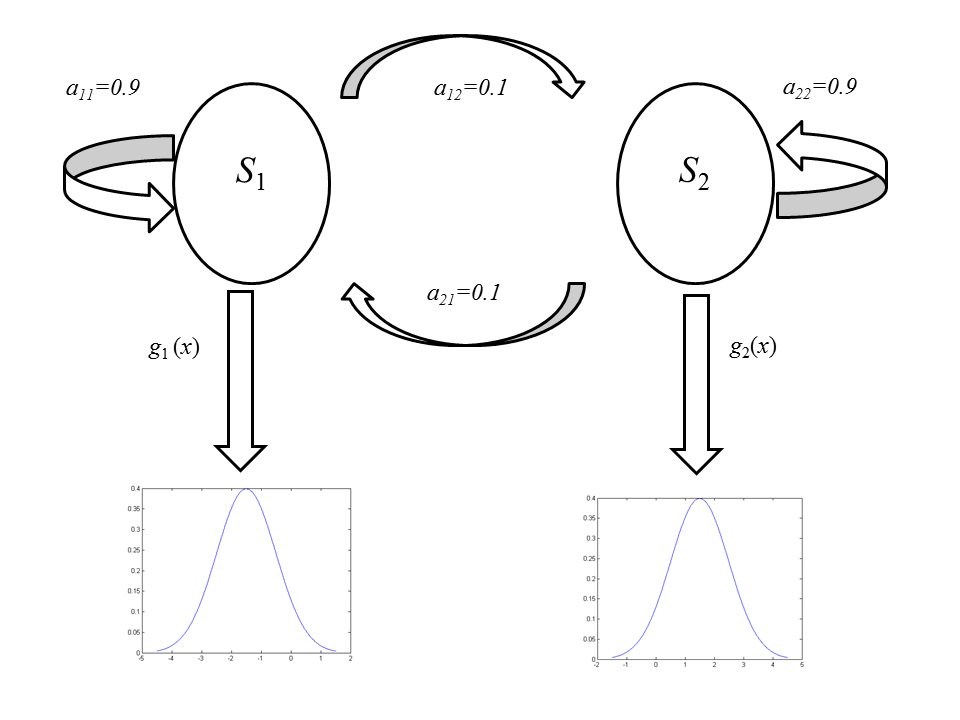}}
	\caption{A hidden Markov source with 2 states, as used in the simulations.}
	\label{fig:HMM}
\end{figure}

Let us assume that  the encoder arbitrarily  uses the last quantizer when it processes the first source symbol.(Any convention can be used as long as it is  known to the decoder. Alternatively we can send an initial side information to determine which quantizer is used.) For the rest of the symbols, the encoder first finds $\underline{\hat{p}}$,  based on which  it selects the suitable quantizer. The exact same procedure is employed by the decoder.

It should be noted that finding $\underline{\hat{p}}$ does not impose a major  computational burden on  the encoder and decoder. They simply update the forward variables (as  discussed earlier) and obtain $\underline{\hat{p}} $. 

A lower bound \cite{switch} on the distortion, when using a  switched scalar quantizer is:
\begin{align}
	\hat{D}= \sum_{m=1}^N \rho_m \{\min_Q \sum_{j=1}^N a_{mj}E_j(x-Q(x))^2\} 
\end{align}

where $E_j$ is expected value using $j$th subsource statistics, $\rho_m$ is the stationary probability of being in state $m$, and minimization is over all possible quantizers. Note that this bound is very optimistic, as it pretends  that the Markov chain state at time $t-1$ is known exactly for selecting the next quantizer. 

\subsection{Results for Synthetic Random Sources}
In the first experiment  the HMS has two Gaussian subsources, one of them having mean $\mu_1=-1.5$ and variance $\sigma^2_1=1$ the other one having mean $\mu_2=+1.5$ and variance $\sigma^2_2=1$. Moreover  $a_{12}=a_{21}=0.1$, as depicted in Fig. \ref{fig:HMM}. For the simulation we set $T=5$ (i.e., the number of codebooks is 5) and the coding rate varies from 1 to 6 bits per symbol. We compare the proposed method with finite state quantizer, as well as DPCM.
\begin{figure*}[t]
	\centering
	\centerline{\includegraphics[width=1\textwidth]{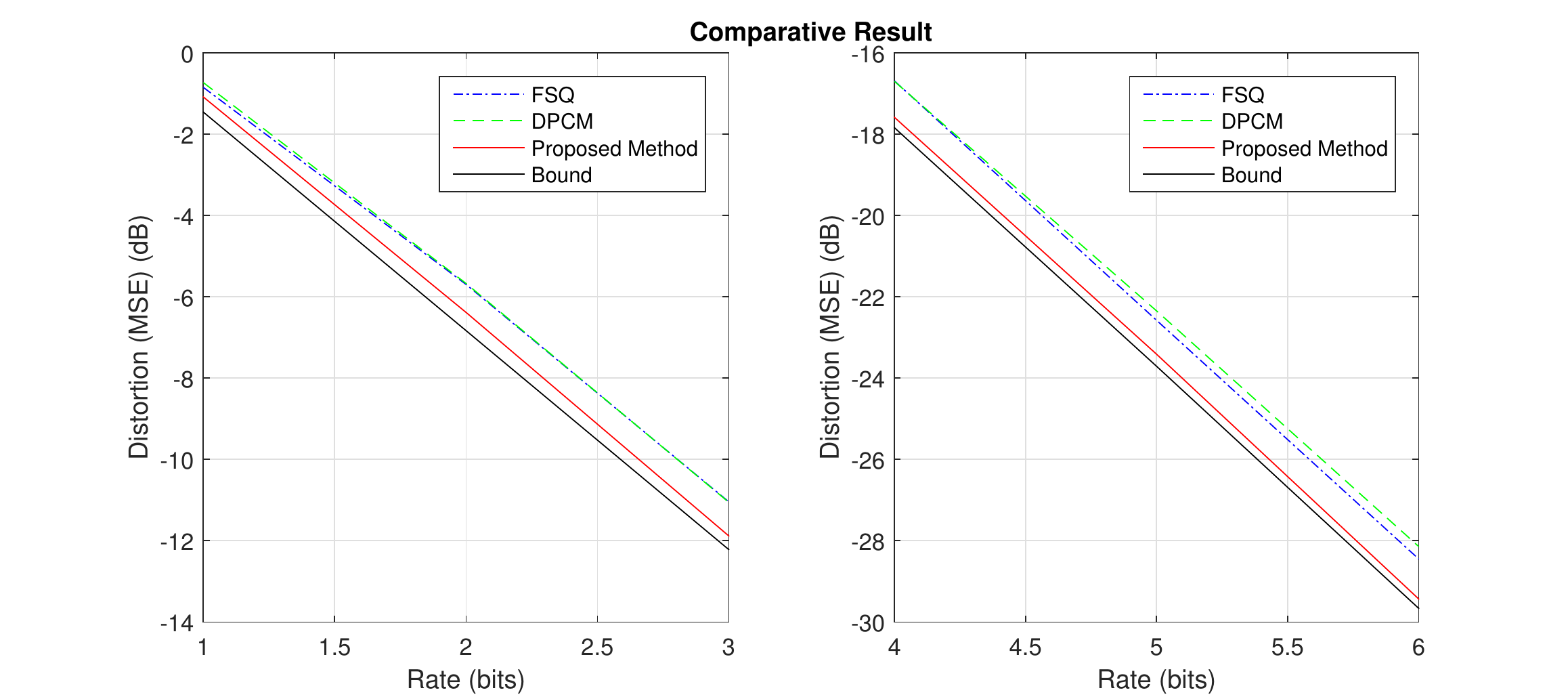}}
	\caption{Rate-Distortion graphs for the proposed method, finite state quantizer (FSQ), DPCM, and the lower bound at bit rates ranging from 1 to 3 and 4 to 6.}
	\label{fig:first}
\end{figure*}
The  result is shown in Fig.\ \ref{fig:first}. As is evident, the proposed method offers gains of approximately 1 dB over the finite state quantizer and DPCM. Comparison to the theoretical lower bound shows that the method leaves little room for improvement and performs very close to the bound.
\begin{figure}[t]
	\centering
	\centerline{\includegraphics[width=.5 \textwidth]{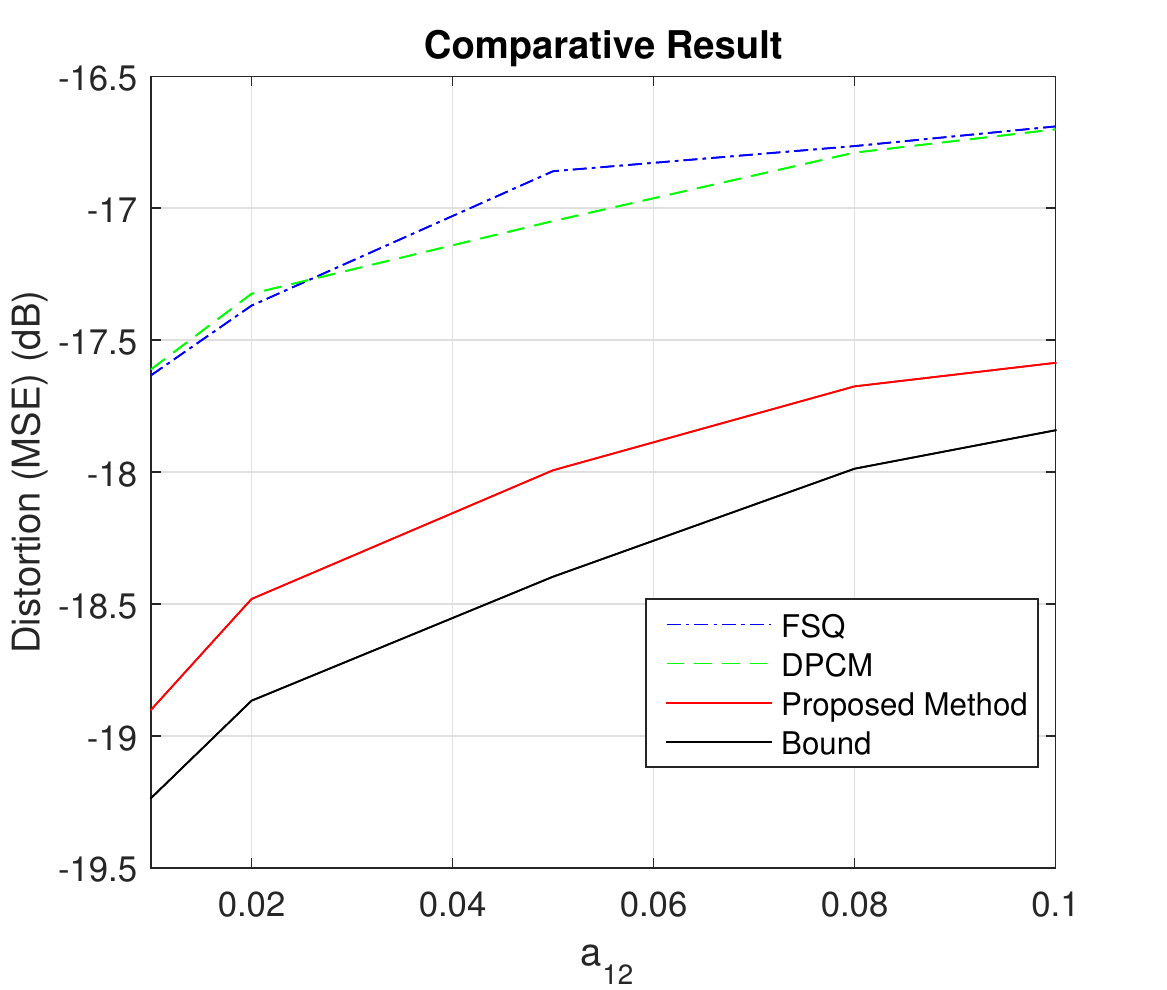}}
	\caption{Rate-Distortion graph for proposed method, finite state quantizer (FSQ), DPCM, and the lower bound  for bit rate 4 bits per symbol for transition probability $a_{12}=a_{21}$ from $0.01$ to $0.1$.  }
	\label{fig:trans}
\end{figure}
Note again that this distortion bound makes the optimistic assumption  that the state at time $t-1$ is known and used to obtain the best encoder for time $t$. Obviously, a real system does  not have access to the source state at any time. So it is not an achievable   lower bound, yet the proposed system performs fairly close to this bound.

The second experiment involves the same source except for varying the transition probability  $a_{12}=a_{21}$ from $ 0.01$ to  $0.1$ with coding rate of 4 bits per symbol.

The method shows consistent gains over its competitors with somewhat larger gains at low values of $a_{12}$ as depicted in Fig. \ref{fig:trans}.

\textbf{Note:} The lower bound introduced earlier, assumes we know the HMM state  at time $t-1$ exactly. However it's a bold assumption and  makes a looser lower bound. A more fair and tighter lower bound could be assuming we know the actual non quantized data till time $t-1$ and based on that we calculate the lower bound. Fig. \ref{newLower} shows the comparison between our proposed approach and the new lower bound. As we could see at higher rate our proposed approach  is almost identical to the lower bound. Since in real design the decoder does not have access to the non quantized data, so the  small gap between the proposed approach and the lower bound at lower rate, does not undermine the optimality of our approach.

\begin{figure}[t]
	\centering
	\centerline{\includegraphics[width=.5 \textwidth]{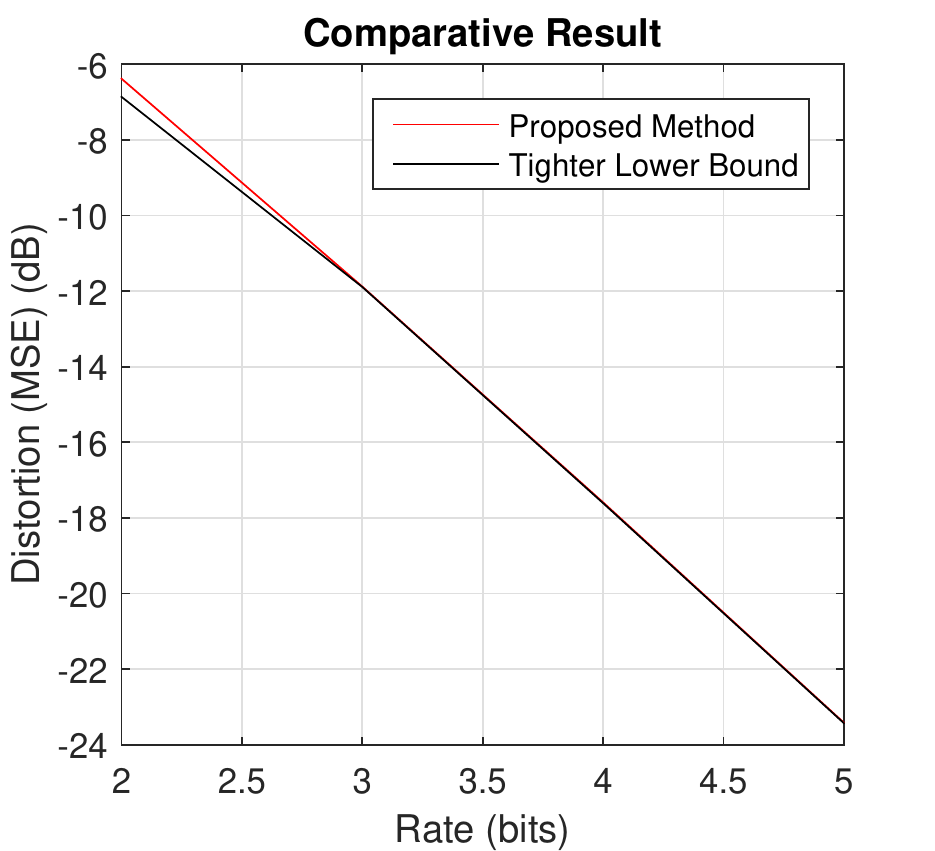}}
	\caption{Rate-Distortion graph for proposed method and the new lower bound  for bit rates ranging from 2 to 5 bits per symbol}
	\label{newLower}
\end{figure}
	
\subsection{Results for the Packet  Loss Case}

As before in the  experiment  the HMS has two Gaussian subsources, one of them having mean $\mu_1=-1.5$ and variance $\sigma^2_1=1$ the other one having mean $\mu_2=+1.5$ and variance $\sigma^2_2=1$. Moreover we ran the experiment on 3 different cases:

Case 1: Transition probability $a_{12}=a_{21}=0.1$

Case 2: Transition probability $a_{12}=a_{21}=0.05$

Case 3: Transition probability $a_{12}=a_{21}=0.01$

\begin{table}[t]
	\centering
	\scalebox{1}{
		\begin{tabular} {|c|c|c|c|}
			\hline
			& FSQ & DPCM &  Proposed Method \\ 
			 \hline
			$No \ Loss$&$-16.61\ dB $& $-16.59 \ dB$ &  $-17.58 \ dB$ \\ \hline
			$1 \% \ Loss$&$-12.65\ dB $& $-12.63 \ dB$ &  $-13.72 \ dB$ \\ \hline
			$5 \% \ Loss$&$-7.31\ dB $& $-7.24 \ dB$ &  $-8.49 \ dB$ \\ \hline
			$10 \% \ Loss$&$-4.61\ dB $& $-4.59 \ dB$ &  $-5.57 \ dB$ \\ \hline				
		\end{tabular}}
		\caption{Distortion for HMM source with transition probability $a_{12}=a_{21}=0.1$ for Lloyd's, DPCM, and Proposed Method, for packet loss $0\ \%$, $1\ \%$, $5\ \%$, and $10\ \%$}
		\label{tab:loss1}
	\end{table}

	\begin{table}[t]
		\centering
		\scalebox{1}{
			\begin{tabular} {|c|c|c|c|}
				\hline
				& FSQ & DPCM &  Proposed Method \\ 
				\hline
				$No \ Loss$&$-16.63\ dB $& $-16.87 \ dB$ &  $-18.05 \ dB$ \\ \hline
				$1 \% \ Loss$&$-12.68\ dB $& $-12.88 \ dB$ &  $-14.08 \ dB$ \\ \hline
				$5 \% \ Loss$&$-7.37\ dB $& $-7.33 \ dB$ &  $-8.60 \ dB$ \\ \hline
				$10 \% \ Loss$&$-4.63\ dB $& $-4.75 \ dB$ &  $-5.57 \ dB$ \\ \hline				
			\end{tabular}}
			\caption{Distortion for HMM source with transition probability $a_{12}=a_{21}=0.05$ for Lloyd's, DPCM, and Proposed Method, for packet loss $0\ \%$, $1\ \%$, $5\ \%$, and $10\ \%$}
			\label{tab:loss2}
		\end{table}

	\begin{table}[t]
		\centering
		\scalebox{1}{
			\begin{tabular} {|c|c|c|c|}
				\hline
				& FSQ & DPCM &  Proposed Method \\ 
				\hline
				$No \ Loss$&$-16.62\ dB $& $-17.64 \ dB$ &  $-19.04 \ dB$ \\ \hline
				$1 \% \ Loss$&$-12.68\ dB $& $-13.15 \ dB$ &  $-14.16 \ dB$ \\ \hline
				$5 \% \ Loss$&$-7.35\ dB $& $-7.49 \ dB$ &  $-8.71 \ dB$ \\ \hline
				$10 \% \ Loss$&$-4.66\ dB $& $-4.84 \ dB$ &  $-5.63 \ dB$ \\ \hline				
			\end{tabular}}
			\caption{Distortion for HMM source with transition probability $a_{12}=a_{21}=0.01$ for LLoyds, DPCM, and Proposed Method, for packet loss $0\ \%$, $1\ \%$, $5\ \%$, and $10\ \%$}
			\label{tab:loss3}
		\end{table}
As is evident in Tab. \ref{tab:loss1},  Tab. \ref{tab:loss2},  and  Tab. \ref{tab:loss3}, for all sources and under different loss scenarios, our proposed approach outperforms the competition. 

\subsection{Results for Real Audio Data}
Linear predictive coding (LPC) is a tool used mostly in audio signal processing and speech processing for representing the spectral envelope of a digital signal of speech in compressed form, using the information of a linear predictive model. It is  one of the most useful methods for encoding good quality speech at a low bit rate and provides extremely accurate estimates of speech parameters. Since LPC is frequently used for transmitting spectral envelope information, and as such it has to be tolerant of transmission errors. Transmission of the filter coefficients directly is undesirable, since they are very sensitive to errors. In other words, a very small error can distort the whole spectrum, or worse, a small error might make the prediction filter unstable. 

Line spectral frequencies (LSF) are used to represent LPC for transmission over a channel. LSFs have several properties (e.g. smaller sensitivity to quantization noise) that make them superior to direct quantization of LPCs.

In our last experiment we used a real speech dataset. We want to design the best coding approach for LSF coefficients. 

Using approaches explained in Section II, we first find the best Guassian emittion HMM, modeling the dataset. Next using the approaches explained in Section III, we design the best codebooks, as well as coding approaches. 

\begin{figure}[t]
	\centering
	\centerline{\includegraphics[width=.5\textwidth]{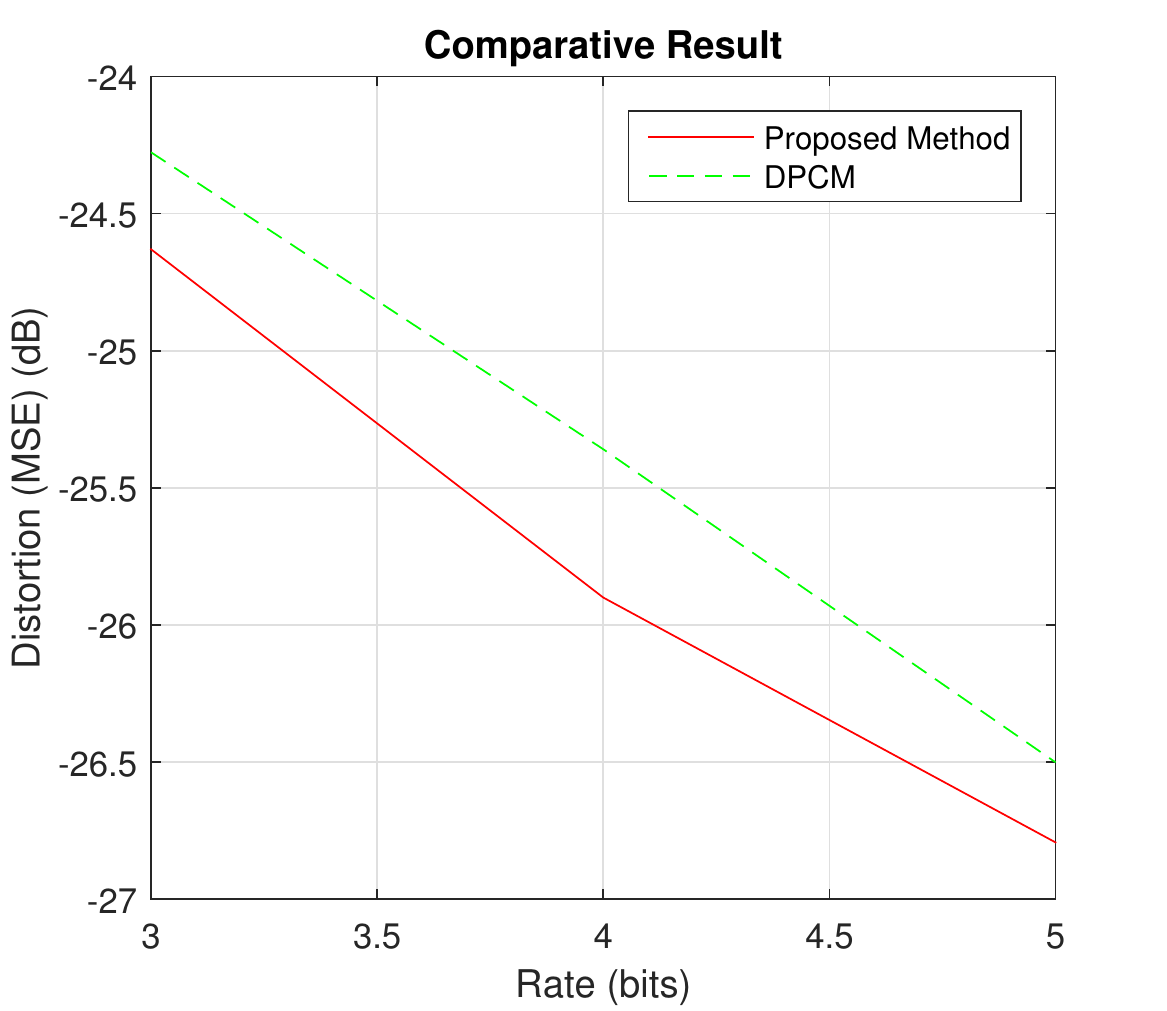}}
	\caption{Rate-Distortion graphs for the proposed method and DPCM at bit rates ranging from 3 to 5.}
	\label{fig:resu}
\end{figure}

As depicted in Fig. \ref{fig:resu}, for coding rates 3 till 5 bits per samples on average we have $0.4 \ dB$ gain compare to DPCM.

\begin{figure}[t]
		\centering
		\centerline{\includegraphics[width=1\linewidth]{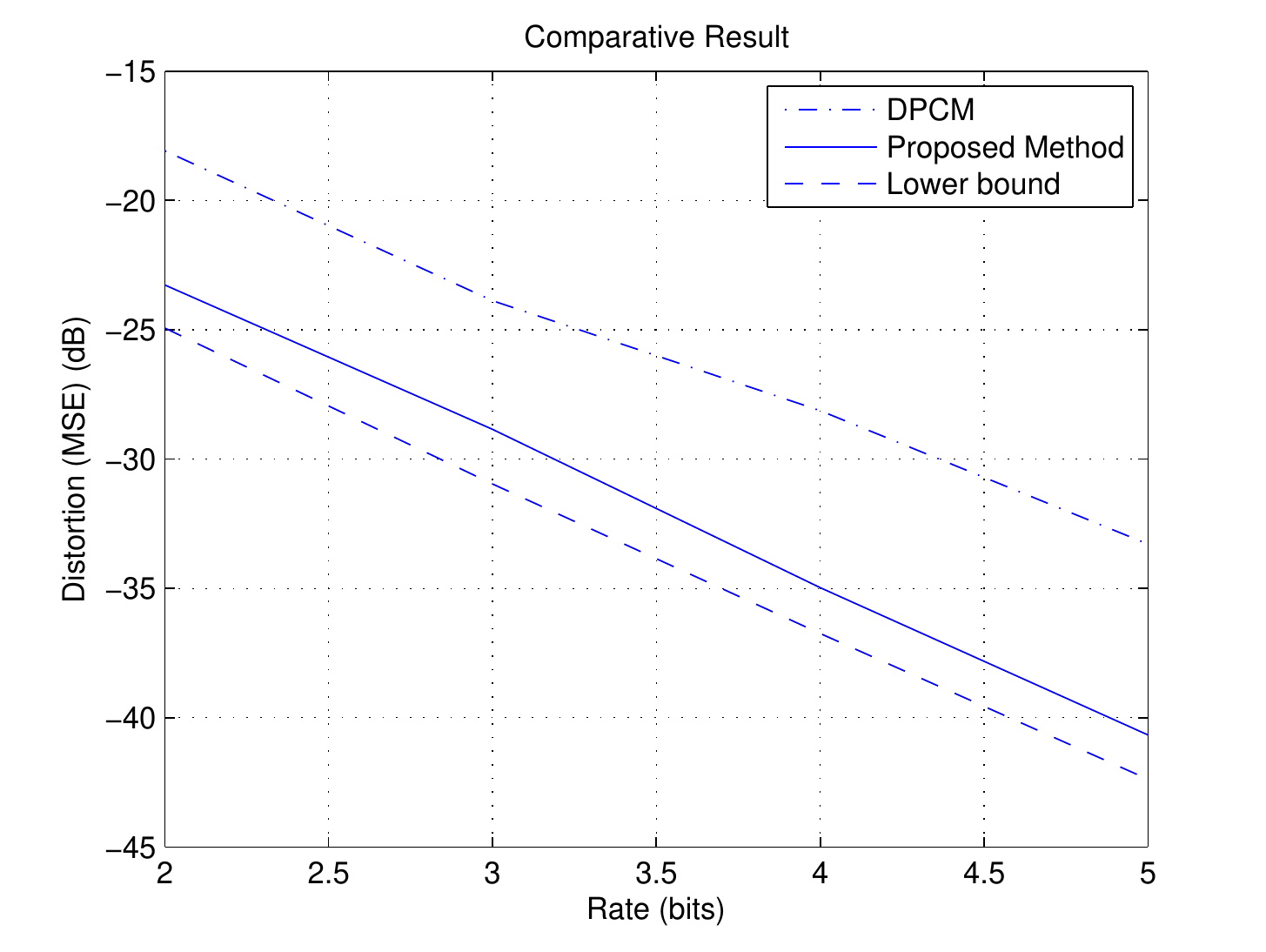}}
		\caption{Rate-Distortion plots for the proposed method, DPCM and the lower bound, at base layer rate of 3 and enhancement layer rate ranging from 2 to 5 bits per sample.}
		\label{fig:result1}
	\end{figure}
\subsection{Results for Standard  Enhancement Layer (Without Access to  Future Base Layer Reconstructions)}
	For  this experiment, we use an HMS which has two Gaussian subsources, one of them with mean $\mu_1=-1.5$ and variance $\sigma^2_1=1$ the other one with mean $\mu_2=+1.5$ and variance $\sigma^2_2=1$, and the transition probabilities are, $a_{12}=a_{21}=0.01$. For the scalable coder, we set the base layer rate to be $R_{12}=3$ bits and the enhancement layer rate varies as $R_2= 2,3,4,5$ bits. We compare the proposed method with the prior approach which assumes a simple Markov model and uses DPCM in the base layer and employs a quantizer designed via Lloyd's algorithm for encoding the base layer reconstruction error in the enhancement layer. Also we compare our results with the theoretical lower bound for using a switched scalar quantizer \cite{switch} operating at the total rate of $ R_{12} + R_2 $ bits. 			
	The rate distortion plots for the different approaches and the lower bound are shown in Fig.\ \ref{fig:result1}. The superiority of the proposed method is evident from the plots with substantial gains of around 5 dB over the competition. Note that there is a gap of around 2 dB from the lower bound due to the scalable coding penalty of the hierarchical structure employed, as this source distortion pair is not successively refinable.

	In the second experiment, the same source is used, but with varying transition probability, $a_{12}=a_{21}$, from $ 0.01$ to  $0.1$, and fixed coding rate of 3 bits in the base and the enhancement layer. Results for this experiment, as shown in Fig. \ref{fig:result2}, demonstrate large gains that decrease marginally with values of $a_{12}$.
	\begin{figure}[t]
		\centering
		\centerline{\includegraphics[width=1\linewidth]{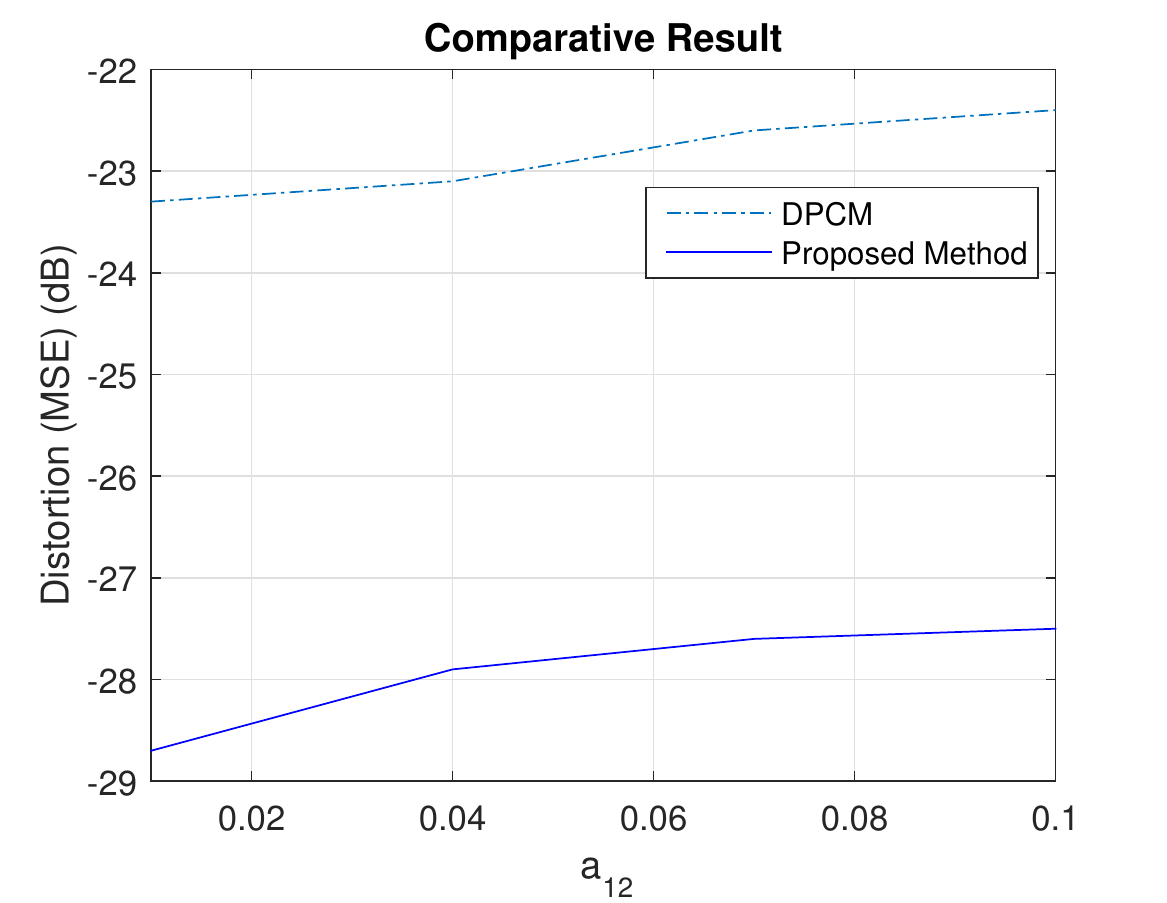}}
		\caption{Distortion plots for the proposed method and DPCM, at base and enhancement layer rates of 3 bits/sample, and transition probability, $a_{12}=a_{21}$, varying from $0.01$ to $0.1$}
		\label{fig:result2}
	\end{figure}
	
	Note that calculating  $\underline{\hat{p}}$ at the base and enhancement layers of both encoder and decoder does not impose  a significant computational burden, as forward variables are easily updated recursively for each sample (as given in Section~\ref{sec:spdtrack}) and then $\underline{\hat{p}} $ is obtained with a few more manipulations.
		\subsection{Results for Enhancement Layer With Access to  Future Base Layer Reconstructions} 
		 We use the same  HMS which has two Gaussian subsources, one of them with mean $\mu_1=-1.5$ and variance $\sigma^2_1=1$ the other one with mean $\mu_2=+1.5$ and variance $\sigma^2_2=1$, and the transition probabilities are, $a_{12}=a_{21}=0.01$. For the scalable coder, we set the base layer rate at $R_{12}=4$ bits and the enhancement layer rate varies at $R_2= 1, 2,3$ bits. We set $L=1$, i.e., using only one future base layer sample.  We measure the additional gains of the proposed method due to access to  future  base layer reconstructed sample. As depicted in the Fig. \ref{fig:result111}, on average the method obtains an additional $.3 \ dB$ gain by exploiting available future  base layer  reconstructed samples.
		 
		 \textbf{Note:} For different values of $L$ we ran the experiment, however the additional gain for $L$ bigger than one was negligible compare to the gain for $L=1$.
		
			\begin{figure}[t]
				\centering
				\centerline{\includegraphics[width=1\linewidth]{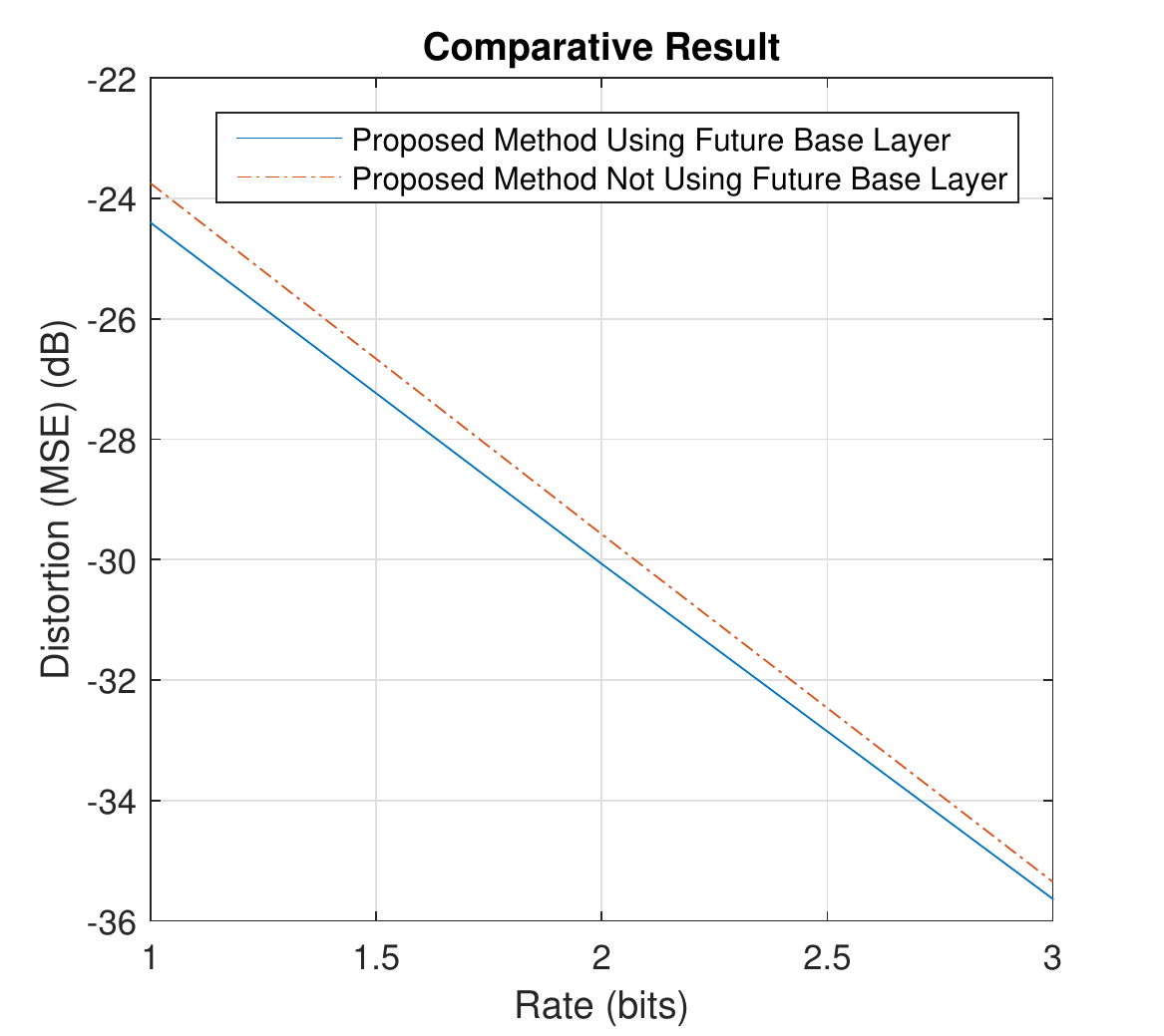}}
				\caption{Rate-Distortion plots for the proposed method with enhancement layer with access to  future base layer reconstructions, and enhancement layer without access to the future base layer reconstructions, at base layer rate of 4 and enhancement layer rate ranging from 1 to 3 bits per sample.}
				\label{fig:result111}
			\end{figure}
			\end{section}
\begin{section}{Conclusion}\label{sec:majhead}
	A new fundamental source coding approach for hidden Markov sources is proposed, based on tracking the state probability distribution, and is shown to be optimal. Practical encoder and decoder schemes that leverage the main concepts are introduced, which consist of three modules: a switchable set of quantizer codebooks, a state probability distribution tracking function, and a next quantizer selector. An iterative approach is developed for optimizing the system given a training set.  

The state probability distribution tracking module estimates, at each time instance, the probability that the source is in each possible state, given the quantized observations, and the next quantizer selector determines the codebook to be used, given the state probability distribution it receives at its input. The decoder mimics the encoder and switches to the same codebook. The switching decisions are made while optimally accounting for all available information from the past, and are obtained efficiently by recursive computation of HMM forward variables.

 Also we have proposed a novel technique for scalable coding of hidden Markov sources which utilizes all the available information while coding a given layer. In the enhancement layer, the state probability distribution is refined for each sample using past enhancement layer reconstructed observations. Then this information is combined with the quantization interval available from the base layer, and the quantizer for the current sample is updated accordingly. The decoder mimics this quantizer updates of the encoder in both the base and enhancement layers.
	
	 Experimental results show substantial performance improvements  over existing approaches. 
	
\end{section}


%

\section*{Acknowledgment}

The authors would like to thank  NSF funding under the code of  NSF-CCF-1016861 and NSF-CCF-1320599.

\ifCLASSOPTIONcaptionsoff
  \newpage
\fi



%

\bibliographystyle{IEEEtran}
\bibliography{dissertation}

%

\begin{IEEEbiography}[{\includegraphics[width=1in,height=1.25in,clip,keepaspectratio]{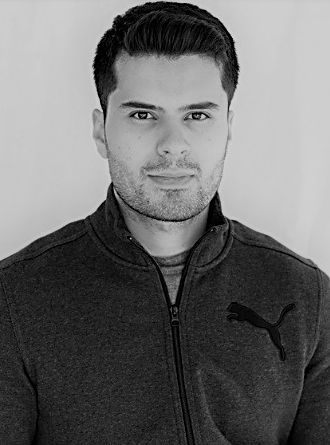}}]{Mehdi Salehifar}
	Mehdi Salehifar (S’17–M’18) received the B.Sc degree in electrical and computer engineering from University of Tehran,Tehran,Iran, in 2012 and the M.Sc. and Ph.D.  degrees in electrical and computer engineering from the University of California, Santa Barbara (UCSB), in 2014 and 2017 respectively. He is currently working in LG Electronics as a senior research engineer. His research interests include signal
	processing, general quantization theory, information theory, and video/audio coding. He worked in LG Electronics as a senior video researcher intern as well as at Qualcomm as an Audio researcher. He is a winner of several titles in national and international mathematic competitions.
\end{IEEEbiography}
\begin{IEEEbiography}[{\includegraphics[width=1in,height=1.25in,clip,keepaspectratio]{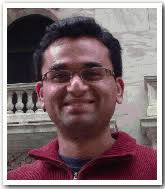}}]{Tejaswi Nanjundaswamy}
	Tejaswi Nanjundaswamy (S’11–M’14) received the B.E degree in electronics and communications engineering from the National Institute of Technology Karnataka, India, in 2004 and the M.S. and Ph.D. degrees in electrical and computer engineering from the University of California, Santa Barbara (UCSB), in 2009 and 2013, respectively. He is currently a post-doctoral researcher at Signal Compression Lab in UCSB, where he focuses on audio/video compression, processing and related technologies. He worked at Ittiam Systems, Bangalore, India from 2004 to 2008 as Senior Engineer on audio codecs and effects development. He also interned in the Multimedia Codecs division of Texas Instruments (TI), India in 2003.Dr. Nanjundaswamy is an associate member of the Audio Engineering Society (AES). He won the Student Technical Paper Award at the AES 129th Convention.
\end{IEEEbiography}
\begin{IEEEbiography}[{\includegraphics[width=1in,height=1.25in,clip,keepaspectratio]{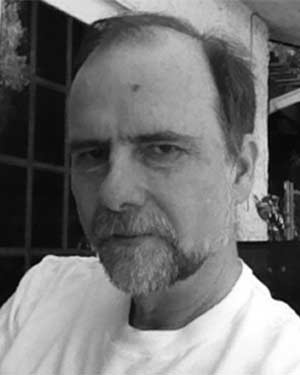}}]{Kenneth Rose}
	Kenneth Rose (S’85–M’91–SM’01–F’03) received the Ph.D. degree from the California Institute of Technology, Pasadena, in 1991.He then joined the Department of Electrical and Computer Engineering, University of California at Santa Barbara, where he is currently a Professor. His main research activities are in the areas of information theory and signal processing, and include rate-distortion theory, source and source-channel coding, audio-video coding and networking, pattern recognition, and non-convex optimization. He is interested in the relations between information theory, estimation theory, and statistical physics, and their potential impact on fundamental and practical problems in diverse disciplines.Prof. Rose was corecipient of the 1990 William R. Bennett Prize Paper Award of the IEEE Communications Society, as well as the 2004 and 2007 IEEE Signal Processing Society Best Paper Awards.
\end{IEEEbiography}




\end{document}